\documentclass[12pt,a4paper]{article}
\usepackage{amsmath,amssymb,textcomp}
\usepackage{graphicx}
\usepackage{dsfont}
\usepackage{braket}
\usepackage{units}

\begin{document}

\title{Perfectly capturing traveling single photons of arbitrary temporal wavepackets with a single tunable device}

\author{Hendra I. Nurdin\thanks{H. I. Nurdin is with the School of Electrical Engineering and Telecommunincations, UNSW Australia, Sydney  NSW 2052, Australia (email: \texttt{h.nurdin@unsw.edu.au}). Research was supported by the Australian Research Council.}, Matthew R. James\thanks{M. R. James is with the School of Engineering, The Australian National University, Canberra ACT 0200, Australia (email: \texttt{matthew.james@anu.edu.au}). Research was supported by the Australian Research Council.}, and Naoki Yamamoto\thanks{N. Yamamoto is with the Department of Applied Physics and Physico-Informatics,
Keio University, Hiyoshi 3-14-1, Kohoku, Yokohama, Japan (email: \texttt{yamamoto@appi.keio.ac.jp}). Research was supported by JSPS Grant-in-Aid No. 15K06151.}}

\date{}

\maketitle

\begin{abstract}
We derive the explicit analytical form of the time-dependent coupling parameter to an external field for perfect absorption of traveling single photon fields  with arbitrary temporal profiles by a tunable single  input-output open quantum system, which can be realized as either a single qubit or single resonator system. However, the time-dependent coupling parameter for perfect absorption has a singularity at $t=0$ and constraints on real systems prohibit a faithful physical realization of the perfect absorber. A numerical example is included to illustrate the absorber's performance under practical limitations on the coupling strength.
\end{abstract}

\section{Introduction}
Quantum networks composed of open quantum systems as network nodes that are interconnected by traveling quantum optical fields are of much interest for quantum information applications, including quantum communication and quantum computing; see, e.g.,  \cite{DLCZ01,Kimb08} and the references therein.  Photons in optical fields are ideal particles for transmitting information between nodes in a quantum network as they can propagate through free space and various physical media. Quantum systems at the nodes are typically matter systems such as an atom, atomic ensembles, superconducting circuits, amongst many possibilities. Each node can  process or store classical or quantum information and exchange its information content with an optical field, either to receive information contained in the field or to transmit information into the field. Thus an important problem for quantum networks is the problem of state transfer between an optical field impinging on a node and the matter system at the node. That is, the transfer of the quantum state of the optical field to the matter and vice-versa.  

The problem of state transfer in a quantum network with two nodes was first considered in the seminal work \cite{CZKM97}. In this work, the authors consider the perfect transfer of an arbitrary superposition state $\alpha |0\rangle + \beta |1\rangle$  from a qubit on one node to a qubit on the other via a one way optical field connecting the two nodes. Each node is a cavity QED system that can be modelled as a qubit coupled to an optical cavity which is in turn coupled to the optical field. It was shown that perfect transfer can be achieved by suitably modulating the Rabi frequency and phase of a Raman laser driving the qubits at each node. They employ the quantum filtering equation or quantum trajectory equation \cite{BvHJ07,HC93} for the network and apply  the dark state principle to derive differential equations characterising the required modulation. The dark state principle states that there should be no photon counted at the output optical field reflected from the receiving node when a photon counting measurement is performed on that output field. 

In this paper, we are interested in the problem of the perfect transfer of a traveling single photon field into a system that can function as a perfect single photon absorber. The state of a travelling single photon field is characterized by a temporal wavepacket $\xi$, a complex-valued function of time, that satisfies $\int_{0}^{\infty} |\xi(s)|^2 ds=1$. The wavepacket gives the detection probability of the single photon state, that is, the probability that the photon will be detected (say, by registering a click in a photo detector) in a time interval $[t_1,t_2]$ ($0 \leq t_1 <t_2$) is given by $\int_{t_1}^{t_2} |\xi(s)|^2 ds$. We are interested in a tunable single photon absorber with a coupling parameter that can be modulated  to absorb single photons of any wavepacket shape.  

In \cite{HRD09} it was shown how single photon fields with symmetric temporal profiles can be mode-matched to be absorbed by a coupled cavity-oscillator system into a single photon state of the oscillator, by suitably modulating the coupling parameter between the cavity and oscillator modes. In the work \cite{DNSK12}, it was shown that an absorber can be implemented with a cavity QED system composed of a three level atom   coupled to an optical cavity. Perfect absorption of a photon with an arbitrary temporal shape could be achieved by appropriately modulating the amplitude of a laser beam driving a Raman transition process in the atom,  under the assumption of no spontaneous emission from the atom. The modulation of the laser beam is tailored to the temporal profile of the incoming single photon field. In this paper, we consider a different class of single photon absorbers, a single two-level system (qubit) or, equivalently, a single resonator with a tunable coupling parameter to an external optical field. This type of coupling has been theoretically proposed  for the transfer of  state between two microwave resonators connected by a one-way travelling optical field  between the resonators \cite{ANK11}. Importantly, such a  tunable coupling parameter has already been demonstrated experimentally in a microwave superconducting resonator \cite{YCSEA13}, where the tunability is realized using an externally controlled variable inductance. This coupling  is analogous to a mirror with tunable transmissivity on an optical cavity. The contribution of this work is to analytically derive the exact form for the time-dependent coupling parameter for perfect absorption of a single photon with an arbitrary wavepacket shape. We describe the system  with a QSDE and derive the optimal evolving coupling parameter by two methods: by explicitly solving the QSDE, and by application of the zero-dynamics principle from \cite{YJ14}. We note that a form of this principle had also been employed in earlier works \cite{HRD09,ANK11,DNSK12} without being referred to as such. That is, these works require that the incoming field and the field reflected from the system interfere destructively, resulting in no photon in the output field (zero output dynamics).

This paper is organized as follows.  Section \ref{sec:preliminaries} introduces the notation of the paper and gives a brief overview of quantum stochastic calculus, quantum stochastic differential equations, single photon generators, and systems driven by a single photon field. Section \ref{sec:tunable-sp-absorber} introduces the model of the single photon absorber of interest, analytically derives the modulating function for the coupling parameter for perfect absorption of any single photon field, and develops a numerical example showing the effects of practical limitations on the absorber's performance. Finally, Section \ref{sec:conclu} draws a conclusion for the paper. 

\section{Preliminaries}
\label{sec:preliminaries}

\noindent \textbf{Notation.} We will use $\imath=\sqrt{-1}$, $^*$ to denote the adjoint of a linear operator
as well as the conjugate of a complex number. If $A=[a_{jk}]$ then $A^{\#}=[a_{jk}^*]$, and $A^{\dag}=(A^{\#})^{\top}$, where $(\cdot)^{\top}$ denotes matrix transposition.  $\Re\{A\}=(A+A^{\#})/2$ and $\Im\{A\}=\frac{1}{2\imath}(A-A^{\#})$.
We use $\mathbb{R}_+$ to denote the set of non-negative real numbers and $L^2(\mathbb{R}_+;\mathbb{C})$ to denote the set of square-integrable functions on $\mathbb{R}_+$.  $\langle X \rangle$ denotes the quantum expectation of an operator $X$, and ${\rm Tr}(X)$ denotes the trace of $X$.

\subsection{Quantum stochastic calculus and quantum stochastic differential equations}
\label{sec:qsc}
We will be working with open Markov quantum systems that are coupled to $n$ continuous-mode boson fields indexed by $j=1,2,\ldots,m$ with annihilation field operators $\eta_j(t)$ satisfying the field commutation relations $[\eta_j(t),\eta_k(t')^*]=\delta_{jk} \delta(t-t')$ and $[\eta_j(t),\eta_k(t')]=0$. For our purposes, we can focus on fields in  a vacuum state. Let us introduce the integrated field annihilation process $A_j(t)=\int_{0}^t \eta_j(s)ds$ and its adjoint process, the integrated field creation process, $A_j^*(t)=\int_{0}^t \eta_j^*(s)ds$. In the vacuum representation, their future-pointing It\={o} increments 
$dA_j(t)=A_j(t+dt)-A_j(t)$ and $dA_j^*(t)=A_j^*(t+dt)-A_j^*(t)$  
satisfy the quantum It\={o} table \cite{HP84,KRP92,Mey95}
\begin{equation*}
\begin{tabular}{l|ll}
$\times $ & $dA_{k}^{*}$ & $dA_{k}$ \\ \hline
$dA_{j}$ & $\delta_{jk} dt$ & 0 \\ 
$dA_{j}^{*}$ & 0 & 0
\end{tabular}
\end{equation*}
We may also define the counting process  (or gauge process)
\begin{equation*}
\Lambda _{jk}(t)=\int_{0}^{t}b_{j}^{\ast }(r)b_{k}(r)dr,
\end{equation*}
which may be included in the It\={o} table \cite{HP84}. The additional
non-trivial products of differentials are
\begin{equation*}
d\Lambda _{jk}dA_{l}^{*}=\delta _{kl}dA_{j}^{*},dA_{j}d\Lambda
_{kl}=\delta _{jk}dA_{l},d\Lambda _{jk}A\Lambda _{li}=\delta _{kl}d\Lambda
_{ji}.
\end{equation*}
Using the processes $A=(A_1,A_2,\ldots,A_m)^{\top}$, $A^{\#}=(A_1^*,A_2^*,\ldots,A_m^*)^{\top}$ and $\Lambda=[\Lambda_{jk}]_{j,k=1,\ldots,m}$, one may define  quantum stochastic integrals of adapted processes on the tensor product of the system and joint boson (symmetric) Fock space of the fields. The system is the quantum mechanical object that is being coupled to the fields, and adapted means that at time $t$ the process acts trivially on the portion of the boson Fock space after time $t$ , see, e.g.,  \cite{HP84,KRP92,Mey95} for details. An adapted process commutes at time $t$ with all of the future pointing differentials.
The product of two adapted processes $X(t)$ and  $Y(t)$ is again adapted, and the increment of the product obeys the quantum It\={o} rule
$$
d(X(t)Y(t)) = (dX(t))Y(t) + X(t) dY(t) + dX(t) dY(t),
$$
Based on these quantum stochastic integrals, one may define quantum stochastic differential equations (QSDEs). 

For the remainder of the paper, we will only consider a system coupled to a single field, $m=1$ and $A(t)=A_1(t)$, $A^*(t)=A_1^*(t)$, $\Lambda(t)=\Lambda_{11}(t)$. An important QSDE that describes the joint unitary evolution of an open Markov system coupled to  vacuum boson fields, common in quantum optics and related fields, is the Hudson-Parthasarathy QSDE given by
\begin{equation}
dU(t) = (-(\imath H(t) +\nicefrac{1}{2}L(t)^*L(t))dt  + dA(t)^*L(t) - L(t)^{*} S(t)dA(t)+(S(t)-I)d\Lambda(t))U(t), \label{eq:HP-QSDE-vac}
\end{equation}
with initial condition $U(0)=I$.   We adopt a very general setting  \cite{BvH08} where $H(t)=H(t)^*$ can be a general adapted process representing the time-dependent Hamiltonian of the system, $L(t)$ an adapted process representing the (possibly time-dependent) coupling of the system to the creation process $A^*(t)$, and $S(t)$  a unitary  adapted process, $S(t)^*S(t)=S(t)S(t)^* = I$, representing the (possibly time-dependent) coupling of the system to the process $\Lambda$ of the field. This QSDE has a unique solution whenever $S(t),L(t),H(t)$ are bounded operators for all $t \geq 0$ \cite{BvH08}. Moreover, in that case the solution is guaranteed to be unitary. The input field $A(t)$ after the interaction is transformed into the output field $Y(t) =U(t)^* A(t) U(t)$. Let $j_t(X)=U(t)^*XU(t)$ denote the Heisenberg picture evolution of a system operator $X$, then we have the following QSDEs for the Heisenberg picture evolution of $j_t(X)$ and $Y(t)$ \cite{GJ09},
\begin{eqnarray*}
dj_t(X) &=& j_t(\mathcal{L}_t(X)) + dA(t)^* j_t(S(t)^*[X,L(t)]) + j_t([L(t)^*,X]S(t))dA(t) \\
&&\quad +  j_t(S(t)^*XS(t)-X) d\Lambda(t),\\
dY(t) &=& j_t(L(t)) + j_t(S(t)) dA(t),
\end{eqnarray*}
with
\begin{eqnarray*}
\mathcal{L}_t(X) = \nicefrac{1}{2} L(t)^*[X,L(t)]+\nicefrac{1}{2}[X,L(t)]L(t)-\imath[X,H(t)].
\end{eqnarray*}
We will denote an input-output open quantum system $G$ with parameters $S(t)$, $L(t)$, $H(t)$ with the notation $G=(S(t),L(t),H(t))$ \cite{GJ09}. The output of system $G_1=(S_1(t),L_1(t),H_1(t))$  can be passed as the input to a second system $G_2=(S_2(t),L_2(t),H_2(t))$ forming a cascaded system which is again an input-output open quantum system $G_3$, the parameters of which is given by the series product rule $\triangleleft$,
$$
G_3 = G_2 \triangleleft G_1 = (S_2(t)S_1(t), L_2(t) + S_2(t) L_1(t), H_1(t) + H_2(t) + \Im\{L_2(t)^* S_2(t) L_1(t)\}).
$$
We note that $G_1$ and $G_2$ can be subsystems of one system sharing the same Hilbert space, see \cite{GJ09} for details. The series product is associative, thus if $G_n =(S_n(t),L_n(t),H_n(t))$ then the series product $G_n \triangleleft G_{n-1} \triangleleft \ldots \triangleleft G_1$ is  unambiguously defined.

\subsection{Single photon generator}
\label{sec:sp-generator}

Our main interest is in systems driven by a traveling single photon field with wavepacket $\xi$, defined by  a continuous-mode single photon state  $| 1_{\xi} \rangle$ on the boson Fock space,  given by $ |1_{\xi} \rangle = \int_{0}^t \xi(s) \eta(s)^* ds\, |\Omega \rangle  = \int_{0}^t \xi(s) dA^*(s)| \Omega \rangle$ \cite{RL00, GJM08,GJN13,GJNC12}, where $|\Omega\rangle$ denotes the vacuum state of the field.

In \cite{GJN11,GJNC12} it has been shown that $|1_{\xi}\rangle$ can be generated as the output of a two-level system (qubit) that is coupled to single vacuum input field through a modulated coupling coefficient. The generator $G_1$ has the Hilbert space $\mathcal{H}_1=\mathbb{C}^2$, where we take as basis vectors $|0 \rangle_1 = (0,1)^{\top}$ and  $|1\rangle_1 = (1,0)^{\top}$, and  is given by $G_1 = (I, \lambda(t) \sigma_{1,-},0)$, where $\sigma_{1,-}$ is the qubit's lowering operator,
$$
\sigma_{1,-} =|0 \rangle_1\langle 1 | =\left[\begin{array}{cc} 0 & 0\\ 1 & 0 \end{array}\right]
$$
and $\lambda(t)$ is a time-dependent complex coupling coefficient of the system to the field given by 
\begin{eqnarray}
\lambda(t) &=& \frac{\xi(t)}{\sqrt{w(t)}}, \label{eq:lambda-gen}
\end{eqnarray}
with $w(t)=\int_{t}^{\infty} |\xi(s)|^2 ds$. The coupling coefficient depends on the wavepacket $\xi$ of the single photon field that one wishes to generate. Let $U_{1}(t)$ denote the unitary solution to the generator's QSDE. When the generator is initialized in the state $|1 \rangle_1$ then we have \cite{GJN11,GJNC12},
$$
U_{1}(t)|1\rangle_1 |\Omega \rangle = \sqrt{w(t)}|1 \rangle_1 \otimes  |\Omega\rangle + |0\rangle_1 \otimes \int_{0}^{t} \xi(s) dA^*(s) |\Omega\rangle
$$
and so $\mathop{\lim}_{t \rightarrow \infty}  U_1(t)|1\rangle | \Omega \rangle =|0 \rangle \otimes  |1_{\xi}\rangle$. Thus, if a photon detection measurement is performed  at the output of the generator at any time $t \geq 0$, the probability of detecting a photon in the interval $[0,t]$ is exactly  $\int_{0}^t |\xi(s)|^2 ds$, as expected.

\subsection{Systems driven by a single photon}
\label{sec:sp-driven}

Systems driven by a traveling single photon field with wavepacket $\xi$ can be viewed equivalently as being driven by an ancillary generator that outputs the single photon field, as described in the preceding subsection. Suppose that a system $G_2=(S_2(t),L_2(t),H_2(t))$ on the Hilbert space $\mathcal{H}_2$ is driven by a single photon field with wavepacket $\xi$. Then the dynamics of the system can be analyzed by studying the {\rm equivalent} cascaded system 
$$
G_2 \triangleleft G_1 = ( S_2(t), L_2(t) +  \lambda(t)S_2(t)^* \sigma_{1,-},H(t)+\Im\{ \lambda(t) L_2(t)^*S_2(t) \sigma_{1,-}\})
$$ 
that is driven by a vacuum field. The cascaded system is defined on the composite Hilbert space $\mathbb{C}^2 \otimes \mathcal{H}_2$.  Note that for notational simplicity for an arbitrary operator $X_1$ on the generator and $X_2$ on the system, we will often denote the ampliations $X_1\otimes I$, $I \otimes X_2$, and $X_1 \otimes X_2$ simply as $X_1$, $X_2$, and $X_1X_2$, etc, with the tensor product being omitted.

\section{A tunable single photon absorber with time-dependent coupling to an external field}
\label{sec:tunable-sp-absorber}

We now describe our tunable photon absorber that can perfectly absorb a single photon field with an arbitrary wavepacket. The Hilbert space of the absorber is that of a qubit, but since only a single excitation will be involved (i.e., a single photon) it can actually be implemented by a quantum harmonic oscillator with a tunable coupling parameter such as the microwave system realized in \cite{DNSK12}. We will say more about the physical realization of the absorber near the end of this section.

We choose $|0 \rangle_2 = (0,1)^{\top}$ and  $|1\rangle_2 = (1,0)^{\top}$ as basis vectors for the absorber Hilbert space $\mathcal{H}_2=\mathbb{C}_2$. Our perfect single photon absorber is then a system $G_2= (I, \gamma(t)\sigma_{2,-},0)$, with  $\sigma_{2,-}$ the qubit lowering operator on $\mathcal{H}_2$. Note that the absorber Hamiltonian has been set to 0, so we are implicitly working in a rotating frame with the respect to the absorber's original Hamiltonian. We only require that the original Hamiltonian commutes with $\sigma_{2,-}$ or only rotates the latter in time as $e^{-\imath \omega t} \sigma_{2,-}$ for some real constant $\omega$ ( $e^{-\imath \omega t}$ can then be absorbed into $\gamma(t)$),  but is otherwise arbitrary. In principle, the analysis to follow can be adapted to accommodate more general Hamiltonians   but we do not pursue this here. The time-dependent parameter $\gamma$ will be chosen to enable perfect absorption of any single photon field, with $\gamma$ dependent on the wavepacket $\xi$ of the latter. Let $G_1$ be the ancillary generator for a single photon as given in Section \ref{sec:sp-generator} for a given $\xi$.  The cascade of $G_1$ onto $G_2$ is given by: 
\begin{equation}
G_2 \triangleleft G_1 = (I,\lambda(t) \sigma_{1,-}  + \gamma(t) \sigma_{2,-}, \Im\{\gamma(t)^* \lambda(t) \sigma_{2,+}\sigma_{1,-} \}), \label{eq:gen-series-abs}
\end{equation}
where $\sigma_{j,+} = \sigma_{j,-}^*$ denotes the raising operator for the system $G_j$. The generator $G_1$ is initialized in the excited state $|1 \rangle_1$ while the absorber $G_2$ is initialized in the ground state $|0 \rangle_2$. The field to which the cascade is coupled to is in the vacuum state $|\Omega\rangle$. We shall show that for any wavepacket $\xi(t)$, $\gamma(t)$ can always be chosen such that in the limit  $t \rightarrow \infty$, the joint system and field  state converges to $|0 \rangle_1 |1 \rangle_2 |\Omega\rangle$. That is, a single photon is asymptotically perfectly absorbed by $G_2$. We shall derive the form for $\gamma$ explicitly in two ways: (i) by explictly solving the QSDE for the cascaded system in Section \ref{sec:gamma-QSDE-solution}, and (ii) by application of the zero-dynamics principle in Section \ref{sec:gamma-zero-dyn}.

It will turn out that the coupling parameter $\gamma$ will have a  singularity at $t=0$ for any wavepacket $\xi$. In actual practical implementation this singularity must be truncated to some finite value. We propose a truncation in Section \ref{sec:absorber-imperfection} and apply it in a numerical example for an exponentially decaying wavepacket to give an illustration of the effect of this truncation on the absorber's performance in capturing a single photon with this wavepacket shape.

Coming back to the discussion at the beginning of this section, we note that since the absorber  can have no more than a single quanta of excitation when driven by a single photon field, the dynamics of $G_2$ when initialized at $|0 \rangle_2$ can be embedded in the dynamics of an equivalent quantum harmonic oscillator system $G_2' = (I, \lambda(t) a, 0)$, where $a$ is the annihilation operator of the oscillator,  satisfying the commutation relation $[a,a^*]=1$. This is because $\mathcal{H}_2$ is isomorphic to ${\rm span}\{|0\rangle_{\rm o},\; |1\rangle_{\rm o} \}$, where $|0 \rangle_{\rm o}$ and $|1 \rangle_{\rm o}$ is the vacuum and 1-photon Fock state of the oscillator, respectively, and $\sigma_{2,-}$ can be identified with the restriction of $a$ to ${\rm span}\{|0\rangle_{\rm o},\; |1\rangle_{\rm o} \}$. Indeed, this kind of embedding of the dynamics of a qubit system into a quantum harmonic oscillator under  single  photon driving has already been exploited in  \cite{PZJ16}.

\subsection{Determining $\gamma$ by solving the QSDE for $G_2 \triangleleft G_1$}
\label{sec:gamma-QSDE-solution}

The wavefunction of the cascaded system (generator and absorber) and the external field is initially $|1 \rangle  |0 \rangle  |\Omega \rangle$ and  at time $t \geq 0$ is of the form $\psi(t)=(\psi_1(t),\psi_2(t),\psi_3(t),\psi_4(t))^{\top} \otimes |\Omega_{[t} \rangle $, where  $\psi(t)$ is adapted on the joint cascaded system  and Fock space, in the sense that  $\psi(t)$ lives in the tensor product of the system and the portion of the Fock space up to time $t$. That is, $U(t) |1 \rangle  |0 \rangle  |\Omega \rangle = \psi(t) \otimes |\Omega_{[t} \rangle$, where $U(t)$ is a solution of the QSDE \eqref{eq:HP-QSDE-vac} with coefficients given by \eqref{eq:gen-series-abs}. Here $|\Omega_{t]} \rangle$ and $|\Omega_{[t} \rangle$ denote the portion of Fock vacuum from time 0 up to time $t$, and from time $t$ onwards, respectively, see \cite{HP84,KRP92}. We have that the vector $\psi(t)$ satisfies the QSDE:
\begin{eqnarray*}
d (\psi(t) \otimes |\Omega_{[t}\rangle) &=& \left[\begin{array}{cccc} -(|\lambda(t)|^2 + |\gamma(t)|^2)/2 & 0 & 0 & 0\\  
0 & -| \lambda(t)|^2/2 & 0 & 0 \\ 0 & -\gamma(t)^*\lambda(t) & -|\gamma(t)|^2/2 & 0 \\ 0 & 0 & 0 & 0 \end{array} \right] \psi(t)  \otimes |\Omega_{[t}\rangle dt\\
&& \quad + \left[\begin{array}{cccc} 0 & 0 & 0 & 0\\  
\gamma(t) & 0 & 0 & 0 \\ \lambda(t) & 0 & 0 & 0\\ 0 & \lambda(t)  &  \gamma(t)  & 0 \end{array} \right] \psi(t) \otimes dA(t)^*|\Omega_{[t}\rangle, 
\end{eqnarray*}
with initial condition $\psi(0) = |1\rangle_1  |0 \rangle_2 = (0,1,0,0)^{\top}$. From the above equation one readily gets that 
\begin{eqnarray*}
\psi_1(t) &=& \exp\left(- \nicefrac{1}{2} \int_{0}^t (|\lambda(s)|^2 + |\gamma(s)|^2) ds \right) \psi_1(0)|\Omega_{t]}\rangle\\
&=& 0,
\end{eqnarray*}
for all $t \geq 0$. Using this fact and recalling from \cite{GJNC12} (using \eqref{eq:lambda-gen}) that 
$$
\exp\left(-\nicefrac{1}{2}\int_{s}^t |\lambda(\tau)|^2 d\tau \right) = \sqrt{\frac{\int_{t}^{\infty} |\xi(\tau)|^2  d\tau}{\int_{s}^{\infty} |\xi(\tau)|^2 d\tau}},
$$
for $0 \leq s \leq t$, we can also solve for $\psi_2(t)$, obtaining
\begin{eqnarray*}
\psi_2(t) &=& \exp\left(- \nicefrac{1}{2}\int_{0}^t |\lambda(\tau)|^2 d\tau \right) \psi_2(0) |\Omega_{t]} \rangle \\
&=& \sqrt{\int_{t}^{\infty} |\xi(\tau)|^2  d\tau} |\Omega_{t]} \rangle.
\end{eqnarray*}
We want that as $t \rightarrow \infty$, $\psi(t) \rightarrow |0 \rangle_1 |1\rangle_2 |\Omega \rangle = (0,0,|\Omega \rangle,0)^{\top}$. That is, $G_2$ completely absorbs the single photon emitted by $G_1$, so that the output field state goes to the vacuum.  

From the equation for $\psi_4(t)$ we have that 
\begin{eqnarray*}
\psi_4(t) &=& \psi_4(0) |\Omega_{t]} \rangle + \int_{0}^t (\lambda(s) \psi_2(s) dA(s)^* + \gamma(t) \psi_3(s) dA(s)^*)|\Omega_{t]}\rangle,\\
&=& \int_{0}^t (\xi(s) dA(s)^* + \gamma(s) \psi_3(s) dA(s)^*)|\Omega_{t]}\rangle,
\end{eqnarray*}
which we want to go to 0 as $t \rightarrow \infty$.  Also, since   
$$
d\psi_3(t) = -\nicefrac{1}{2}|\gamma(t)|^2  \psi_3(t)dt -  \gamma(t)^* \lambda(t) \psi_2(t) dt,
$$ 
we can solve this to give:
\begin{eqnarray*}
\psi_3(t) &=& \exp\left(-\nicefrac{1}{2}\int_{0}^t |\gamma(s)|^2 ds \right) \psi_3(0) |\Omega_{t]}\rangle \\
&&\quad - \int_{0}^t \exp\left(-\nicefrac{1}{2}\int_{s}^t |\gamma(\tau)|^2 d\tau \right)
\gamma(s)^*\lambda(s)\psi_2(s) ds)|\Omega_{t]}\rangle,\\
 &=&  - \int_{0}^t \exp\left(-\nicefrac{1}{2}\int_{s}^t |\gamma(\tau)|^2 d\tau \right)
\gamma(s)^*\lambda(s)\psi_2(s) ds)|\Omega_{t]}\rangle,\\
\end{eqnarray*}
the second line following from the fact that $\psi_3(0)=0$. The expressions for $\psi_4(t)$ and $\psi_3(t)$ just given, and  the form of $\lambda(t)$ for the generator, suggest the following as an educated guess for $\gamma(t)$ to achieve $\mathop{\lim}_{t \rightarrow \infty} \psi_4(t)=0$ and $\mathop{\lim}_{t \rightarrow \infty} \psi_3(t)=| \Omega \rangle$, 
$$
\gamma(t) = -e^{\imath \phi_0} \frac{\xi(t)}{\sqrt{\int_{0}^t |\xi(s)|^2 ds}},
$$ 
for some arbitrary real constant $\phi_0$. We will verify below that this form of the modulation function $\gamma(t)$ indeed achieves perfect absorption. Note that although $\gamma$ is singular at $t=0$, $\gamma(0)=-\infty$, it is square integrable on $[s,t]$ for any $0 < s \leq t \leq \infty$. Indeed, by direct integration,
\begin{eqnarray*}
\int_{s}^t |\gamma(\tau)|^2 d\tau &=& \int_{s}^t \left(|\xi(\tau)|^2/\int_{0}^{\tau} |\xi(y)|^2 dy \right) d\tau,\\
&=& \int_{s}^t d \left(\ln \int_{0}^{\tau} |\xi(y)|^2dy\right),\\
&=& \ln \left(\frac{\int_{0}^{t} |\xi(y)|^2 dy}{\int_{0}^{s} |\xi(y)|^2 dy} \right). 
\end{eqnarray*}
Moreover,  from this it follows immediately that
\begin{eqnarray*}
\exp\left(-\nicefrac{1}{2}\int_{s}^t |\gamma(\tau)|^2 d\tau \right) &=& \left\{\begin{array}{cc} \sqrt{\frac{\int_{0}^{s} |\xi(y)|^2 dy}{\int_{0}^{t} |\xi(y)|^2 dy}} & s>0 \\  0 & s = 0,t>0 \\ 1 & s=t=0 \end{array}\right.,
\end{eqnarray*}
which is thus well-defined for any $0 \leq s \leq t \leq \infty$.  Substituting the expression for $\gamma(t)$ and solving for $\psi_3(t)$ gives 
\begin{eqnarray*}
\psi_3(t) &=& \left\{ \begin{array}{cc} e^{-\imath \phi_0}\frac{1}{\sqrt{\int_{0}^{t} |\xi(\tau)|^2 d\tau}} \int_{0}^t |\xi(s)|^2 ds | \Omega_{t]} \rangle & t>0 \\ 0 & t=0 \end{array}. \right.
\end{eqnarray*}
Notice that $\mathop{\lim}_{t \rightarrow \infty} \psi_3(t) =  | \Omega \rangle$, as desired. Substituting all of the above into the expression for $\psi_4(t)$ gives,
\begin{eqnarray*}
\psi_4(t) &=& \int_{0}^t (\lambda(s) \psi_2(s) dA(s)^* + \gamma(s) \psi_3(s) dA(s)^*)|\Omega_{t]}\rangle,\\
&=& \int_{0}^t (\xi(s) dA(s)^* - \xi(s) dA(s)^*)|\Omega_{t]} \rangle,\\
&=& 0,
\end{eqnarray*}
again exactly as expected.  Thus, we conclude that a two-level system of the form $G =( I, \gamma(t) \sigma_{-},0)$ with $\gamma(t) = -e^{\imath \phi_0}\xi(t)/\sqrt{\int_{0}^{t} |\xi(\tau)|^2 d\tau}$ can perfectly absorb a single photon with an arbitrary wavepacket $\xi(t)$.

\subsection{Determining $\gamma$ by application of the zero-dynamics principle}
\label{sec:gamma-zero-dyn} 
 
We now show that the modulation signal $\gamma(t)$ can also be found by application of the zero-dynamics principle proposed in \cite{YJ14}. 
Recall that the  cascaded generator and absorber has the coupling operator $L(t) =\lambda(t) \sigma_{1,-} + \gamma(t) \sigma_{2,-}$ and the Hamiltonian is $H(t) =  \Im\{\gamma(t)^* \lambda(t) \sigma_{2,+}\sigma_{1,-} \}$. 
The initial state of the composite generator and absorber system is  $|1\rangle_1 |0 \rangle_2$. The formulation of the zero-dynamics principle given in \cite{YJ14} is in the Heisenberg picture. It states that the output field $Y(t)= \int_{0}^t j_s(L(s))ds +A(t)$ should be a vacuum field. This is necessary as no photon should be present in the output field if perfect absorption is taking place.
This means that one must have \cite{GJN10},
$$
\left\langle \exp\left(\imath \int_{0}^{\infty} (u(t)^{*}dY(t) + u(t)  dY(t)^{*})\right) \right\rangle = \exp\left( - \nicefrac{1}{2} \int_{0}^{\infty} |u(s)|^2 ds \right),
$$
for any $u \in L^2(\mathbb{R}_+;\mathbb{C})$, where the right hand side is the characteristic function of the vacuum state of a bosonic field. Evaluating the left hand side is in general difficult, if not intractable, except in special instances like the linear quantum networks analysed in \cite{YJ14}. However, the zero-dynamics principle can still be applied, but by looking at it from another equivalent viewpoint. Zero-dynamics corresponds to the state of the output field always being vacuum at all times $t \geq 0$, $U(t) |1\rangle_1 |0 \rangle_2 |\Omega\rangle = |\varphi(t) \rangle |\Omega\rangle$ for some pure state vector $|\varphi(t)\rangle$ on $\mathcal{H}_1 \otimes \mathcal{H}_2$ for all $t \geq 0$. This is similar in spirit to the form of the principle employed in \cite{ANK11,DNSK12}.  Thus we may state the ansatz that the wavefunction is $|\varphi(t) \rangle = \alpha(t) |1\rangle_1 |0 \rangle_2 + \beta(t) |0\rangle_1 |1 \rangle_2$ for $t \geq 0$ for some scalar complex functions $\alpha(t)$ and $\beta(t)$  that will be sought, satisfying $|\alpha(t)|^2 + |\beta(t)|^2=1$. We also have the initial condition $\alpha_0 = 1$ and $\beta_0=0$. Substituting this ansatz into the joint QSDE of the system and field gives:
$$
d|\varphi(t) \rangle \otimes |\Omega\rangle=\left(-(\imath H(t) +\nicefrac{1}{2}L(t)^*L(t))dt + L(t) dA(t)^* - L(t)^* dA(t) \right) |\varphi(t) \rangle \otimes |\Omega\rangle.
$$
Since the input to the generator-absorber is vacuum, for zero-dynamics to hold the term $L(t)dA(t)^*$ in the QSDE should be 0 for all times (otherwise $dA^*(t)$ will create a photon in the field after time $t$ and the field will no longer be in the  vacuum state), meaning that $L(t) |\varphi(t) \rangle = 0$ for all $t \geq 0$. This gives the condition
\begin{equation*}
(\lambda(t)  \alpha(t) +\beta(t) \gamma(t))|0 \rangle_1 |0 \rangle_2 =0,\;\forall t\geq 0 
\end{equation*}
and thus,
\begin{equation}
\lambda(t)  \alpha(t) +\beta(t) \gamma(t)=0, \; \forall t \geq 0. \label{eq:albe-cons}
\end{equation}
Moreover, noting that $L(t)^*dA(t) |\varphi(t) \rangle |\Omega\rangle=0$ (since $dA(t)$ annihilates the portion of the vacuum after time $t$), the QSDE reduces to the deterministic equations,
\begin{eqnarray*}
d|\varphi(t) \rangle &=& -\imath H(t)  |\varphi(t) \rangle dt,\\
\dot{\alpha}(t) |1\rangle_1 |0\rangle_2 + \dot{\beta}(t) |0\rangle_1 |1\rangle_2 &=&  \nicefrac{1}{2}\gamma(t)\lambda(t)^*\beta(t) |1\rangle_1 |0\rangle_2  -\nicefrac{1}{2}\gamma(t)^* \lambda(t)\alpha(t)|0\rangle_1 |1\rangle_2
\end{eqnarray*}
Thus we arrive at the following set of differential equations,
\begin{eqnarray*}
\dot{\alpha}(t) &=& \gamma(t) \lambda(t)^* \beta(t),\\
\dot{\beta}(t) &=& -\gamma(t)^* \lambda(t)\alpha(t),\\
\lambda(t)  \alpha(t) +\beta(t) \gamma(t) &=& 0
\end{eqnarray*}
Using the constraint (\ref{eq:albe-cons}), subtituting $\beta(t) \gamma(t) = -\lambda(t) \alpha(t)$ gives,
\begin{eqnarray*}
\dot{\alpha}(t) &=& -\nicefrac{1}{2} |\lambda(t)|^2 \alpha(t),\\
\dot{\beta}(t) &=& \nicefrac{1}{2}|\gamma(t)|^2 \beta(t),\\
\lambda(t)  \alpha(t) +\beta(t) \gamma(t) &=& 0
\end{eqnarray*}
The solution for $\alpha$ with initial condition $\alpha_0=1$ is:
\begin{eqnarray*}
\alpha(t) &=& e^{-\int_{0}^t |\lambda(s)|^2 ds} \alpha_0 = \sqrt{\int_{t}^{\infty} |\xi(s)|^2 ds},\\
\end{eqnarray*}
But we also have that $|\alpha(t)|^2 + |\beta(t)|^2 =1$, so $|\beta(t)|^2 = \int_{0}^{t} |\xi(s)|^2ds$. It follows from the constraint \eqref{eq:albe-cons} that  
$$
|\gamma(t)| = |\lambda(t) \alpha(t)/\beta(t)| = \frac{|\xi(t)|}{\sqrt{\int_{0}^{t} |\xi(s)|^2ds}}.
$$ 
So, we arrive at
\begin{eqnarray*}
\beta(t) &=& \sqrt{\int_{0}^t |\xi(s)|^2ds}\, e^{-i\phi(t)},\; \gamma(t) =  -\frac{\xi(t)}{\sqrt{\int_{0}^t |\xi(s)|^2ds}} e^{i \phi(t)}. 
\end{eqnarray*}
for some arbitrary real-valued differentiable function $\phi(t)$. However, to satisfy the differential equation for $\beta(t)$, $\phi(t)$ must in fact be a real constant, say, $\phi_0$. Whence,
\begin{eqnarray*}
\beta(t) &=& \sqrt{\int_{0}^t |\xi(s)|^2ds}\, e^{-i\phi_0},\;  \gamma(t) = - \frac{\xi(t)}{\sqrt{\int_{0}^t |\xi(s)|^2ds}} e^{i \phi_0}. 
\end{eqnarray*}
Thus we recover the form of the modulating function $\gamma(t)$ that was obtained in Section \ref{sec:gamma-QSDE-solution} by explicitly solving the QSDE, indicating the power of the zero-dynamics principle for state-transfer problems such as this. Since one does not need to solve a QSDE, this approach would be more widely applicable. Moreover, even if the QSDE is explicitly solvable, it allows the differential equations characterizing $\gamma(t)$ to be derived thus avoiding having to make an educated guess about $\gamma(t)$. 

\subsection{Effect of imperfect realization of $\gamma$}
\label{sec:absorber-imperfection}

We have seen that the analytical form of $\gamma$ for perfect absorption has a singularity that grows as $O(t^{-\nicefrac{1}{2}})$ as $t \rightarrow 0$ for any wavepacket $\xi$. This poses a practical challenge  as  an infinite coupling magnitude cannot be physically realized. Moreover, in the potential implementation of the absorber using a single mode resonator, the coupling parameter should be much smaller than the resonator's free spectral range as to not excite higher frequency modes of the resonator. A sub-optimal implementation would be to truncate the magnitude of $\gamma(t)$ for small values of $t$ to a finite value. However, this will mean that the absorber will no longer perfectly absorb a single photon. In this section, we numerically evaluate the effect of bounding the magnitude of $\gamma(t)$ for small $t$ on a particular example.

Let   $n_j =\sigma_{j,+} \sigma_{j,-}$ denote the number operator for the generator ($j=1$) and absorber ($j=2$). To ease the notation, in the following we will often not explicitly write the time dependence on operators, with the time dependence being implicitly understood. We will be interested in the evolution of the mean number operator for the absorber, $\langle n_2 \rangle$, in the Heisenberg picture, as this gives the probability of excitation of a single photon in the absorber. In the perfect absorption case we have already discussed, we have that $\mathop{\lim}_{t \rightarrow \infty} \langle n_2 \rangle = 1$.   Since the combined generator-absorber is driven by a vacuum field, the evolution of the mean $\langle X_1 X_2 \rangle$ is given by the ordinary differential equation (the general expression for $\mathcal{L}_t(X_1X_2) $ is derived in the Appendix),
\begin{eqnarray*}
\frac{d}{dt} \langle X_1X_2 \rangle &=&  \langle \mathcal{L}_t(X_1X_2) \rangle,\\
&=& \gamma(t)^* \lambda(t) \bigl \langle X_1 \sigma_{1,-} [\sigma_{2,+},X_2] \bigr \rangle \\
&& +\nicefrac{1}{2}  |\lambda(t)|^2\bigl \langle (\sigma_{1,+}[X_1,\sigma_{1,-}]+[\sigma_{1,+},X_1]\sigma_{1,-})X_2 \bigr \rangle \\
&& \quad +\lambda(t)^* \gamma(t) \bigl \langle \sigma_{1,+} X_1[X_2,\sigma_{2,-}] \bigr \rangle \\
&& +  \nicefrac{1}{2} |\gamma(t)|^2 \bigl \langle (\sigma_{2,+}[X_2,\sigma_{2,-}]+[\sigma_{2,+},X_2]\sigma_{2,-})X_1 \bigr\rangle.
\end{eqnarray*}
When $X_2=n_2 $ and $X_1=1$, we get that:
\begin{eqnarray}
\frac{d}{dt} \langle n_2 \rangle &=&  \langle \mathcal{L}_t(n_2) \rangle,\\
&=& \gamma(t)^* \lambda(t) \bigl \langle \sigma_{1,-} [\sigma_{2,+},n_2] \bigr \rangle +\lambda(t)^* \gamma(t) \bigl \langle \sigma_{1,+}[n_2,\sigma_{2,-}] \bigr \rangle \notag\\
&& \qquad +\nicefrac{1}{2}|\gamma(t)|^2 \bigl \langle (\sigma_{2,+}[n_2,\sigma_{2,-}]+[\sigma_{2,+},n_2]\sigma_{2,-})\bigr\rangle \notag,\\
&=& -|\gamma(t)|^2 \langle n_2 \rangle -\gamma(t)^* \lambda(t) \bigl \langle \sigma_{1,+} \sigma_{2,-} \bigr\rangle^*  - \gamma(t) \lambda(t)^* \bigl \langle \sigma_{1,+} \sigma_{2,-} \bigr\rangle \label{eq:mean-eq-1}.
\end{eqnarray}
Let $\sigma_{2,z} = [\sigma_{2,+},\sigma_{2,-}]$. Then when $X_2=\sigma_{2,-}$ and $X_1=\sigma_{1,+}$ the equation is:
\begin{eqnarray}
\frac{d}{dt} \langle \sigma_{1,+} \sigma_{2,-} \rangle &=&  \langle \mathcal{L}(\sigma_{1,+}\sigma_{2,-}) \rangle, \notag \\
&=&-\nicefrac{1}{2}(|\lambda(t)|^2 +|\gamma(t)|^2)\langle \sigma_{1,+} \sigma_{2,-}\rangle +  \gamma(t)^* \lambda(t) \bigl \langle n_1 \sigma_{2,z} \rangle. \label{eq:mean-eq-2}
\end{eqnarray}
When $X_1 = n_1$ and $X_2 = \sigma_{2,z}$, the equation is:
\begin{eqnarray}
\frac{d}{dt} \langle n_1 \sigma_{2,z} \rangle &=&  \langle \mathcal{L}_t(n_1\sigma_{2,z}) \rangle, \notag\\
&=&-|\lambda(t)|^2 \langle n_1 \sigma_{2,z}\rangle +  2 |\gamma(t)|^2 \bigl \langle n_1 n_2\rangle. \label{eq:mean-eq-3}
\end{eqnarray}
Finally, when $X_1= n_1$ and $X_2=n_2$ the equation is:
\begin{eqnarray}
\frac{d}{dt} \langle n_1 n_2 \rangle &=&  \langle \mathcal{L}_t(n_1n_2) \rangle, \notag\\
&=&-(|\lambda(t)|^2  + |\gamma(t)|^2) \langle n_1 n_2 \rangle \label{eq:mean-eq-4}.
\end{eqnarray}
The initial conditions are $\langle n_2(0) \rangle =0$, $\langle \sigma_{1,+}(0)\sigma_{2,-}(0)\rangle = 0$, $\langle n_1(0) \sigma_{2,z}(0) \rangle =-1$, and $\langle n_1(0)n_2(0) \rangle=0$. From the last initial condition, we can solve for \eqref{eq:mean-eq-4} to get that $\langle n_1(t)n_2(t) \rangle =0$ for all $t \geq 0$. Thus there are actually only three coupled equations that can have a non-trivial solution. The remaining initial conditions can then be used to sequentially solve equations \eqref{eq:mean-eq-3}, \eqref{eq:mean-eq-2},  and \eqref{eq:mean-eq-1} (in that order), yielding the explicit solutions. However, here we are interested in evaluating the effect of approximating the function $\gamma(t)$, which is singular at $t=0$, with a function $\gamma_{T}(t)$ given by
$$
\gamma_T(t) = \left\{ \begin{array}{cc}  - \frac{\xi(T)}{\sqrt{\int_{0}^T |\xi(s)|^2 ds}}, &  t \leq T, \\
- \frac{\xi(t)}{\sqrt{\int_{0}^t |\xi(s)|^2 ds}}, & t > T \end{array} \right.,
$$
where $T$ is a real parameter taking a value in the open interval $(0,\infty)$. By construction, $\gamma_T(t)$ is continuous for all $t \geq 0$. To evaluate the probability of the absorber being excited to the single photon state $|1\rangle _2$ when $\gamma(t)$ is replaced by $\gamma_T(t)$ for some chosen value of $T$, we can solve for \eqref{eq:mean-eq-3} to obtain
$$
\langle n_1(t)\sigma_{2,z}(t) \rangle = -  \int_{t}^{\infty} |\xi(\tau)|^2 d\tau,
$$
and numerically integrate Eqs. \eqref{eq:mean-eq-1} and \eqref{eq:mean-eq-2}, with the given initial conditions. 

\begin{figure}[!t]
\centering
\includegraphics[width=0.8\textwidth]{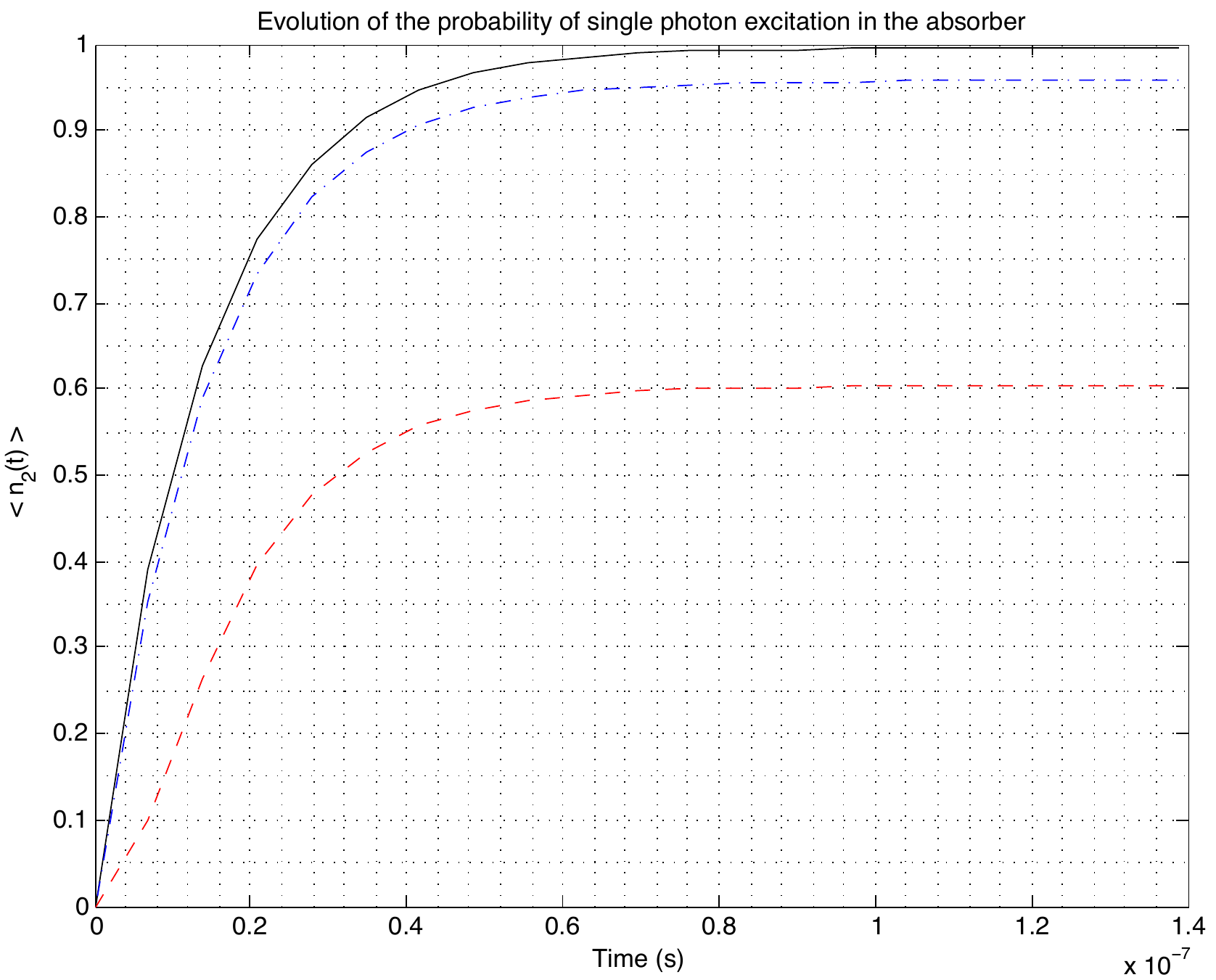}
\caption{Time evolution of the excitation probability $\langle n_2 \rangle$. with the absorber initalized in the state $|0 \rangle_2$, from $t=0$ to $t=t_1=1.3889 \times 10^{-7}$, for $T=0.001t_1$ (top solid black curve), $T=0.01t_1$ (middle dashed blue curve), and $T=0.1t_1$ (bottom dash dotted red curve).}
\label{fig:absorber-plots}
\end{figure}

Let us consider the case where the input wavepacket  $\xi(t)$ is a decaying exponential function of the form
$
\xi(t) = \sqrt{c}e^{-ct/2}$
for some positive real constant $c$. This is the form of the wavepacket that would be produced at the output of an optical cavity that is  initialized in the 1-photon Fock state. Let us take $c = 7.2 \times 10^7$ (this is a value that can be realized in table-top quantum optical experiments, see, e.g., \cite{IYYYF11}) and allow the absorber to run up to time $t_1 = 10/c = 1.3889 \times 10^{-7}$. Fig.~\ref{fig:absorber-plots} shows the evolution of $\langle n_2(t) \rangle$ for $T= 0.001t_1$, $T  =0.01t_1$, and $T =0.1t_1$. It can be seen that for larger $T$ (wider truncation) the excitation probabillity of the absorber is lower for all  $t \geq 0$, as can be expected. The steady-state excitation probability is approximately  0.9957, 0.9575, and 0.6037 for $T= 0.001t_1$, $T =0.01t_1$, and $T =0.1t_1$, respectively.

%

\section{Conclusion}
\label{sec:conclu}
In this work we have considered a single photon absorber with a tunable coupling parameter to an external travelling single photon field. We analytically derived the exact form of the time-dependent coupling parameter for perfect absorption of a single photon field of any temporal wavepacket shape. The ideal modulation function has a singularity at $t=0$ which cannot be attained in real devices, therefore it is  approximated with a continuous function that is truncated to a finite value at $t=0$. In a numerical example, we illustrate the effect of this truncation on the ability of the absorber to perfectly absorb a single photon for a particular truncation scheme applied to an exponentially decaying wavepacket.

%


\appendix
\section*{Appendix: General expression for $\mathcal{L}_t(X_1X_2)$}

\begin{eqnarray*}
\mathcal{L}_{L(t)}(X_1 X_2) &=&\nicefrac{1}{2}L(t)^*[X_1X_2,L(t)]+ \nicefrac{1}{2}[L(t)^*,X_1X_2]L(t),\\
&=& \nicefrac{1}{2} (\lambda(t) \sigma_{1,+}+\gamma(t)^* \sigma_{2,+})[X_1 X_2, \lambda(t) \sigma_{1,-}+\gamma(t) \sigma_{2,-}]\\
&&\quad +\nicefrac{1}{2}[\lambda(t)^* \sigma_{1,+}+\gamma(t)^* \sigma_{2,+},X_1X_2](\lambda(t) \sigma_{1,-}+\gamma(t) \sigma_{2,-})\\
&=& \nicefrac{1}{2} (\lambda(t)^* \sigma_{1,+}+\gamma(t)^* \sigma_{2,+})(\lambda(t)[X_1,\sigma_{1,-}]X_2+\gamma(t) X_1 [X_2,\sigma_{2,-}])\\
&&\quad +\nicefrac{1}{2}(\lambda(t)^* [\sigma_{1,+},X_1]+\gamma(t)^* \sigma_{2,+}X_1[\sigma_{2,+},X_2](\lambda(t) \sigma_{1,-}+\gamma(t) \sigma_{2,-}).\\
-\imath [X_1X_2,H(t) ]&=& -\imath [X_1 X_2,\Im\{\gamma(t)^*\lambda(t) \sigma_{2,+} \sigma_{1,-}\}],\\
&=& -\imath[X_1 X_2, \nicefrac{1}{2\imath}(\gamma(t)^* \lambda(t) \sigma_{2,+} \sigma_{1,-} - \gamma(t) \lambda(t)^* \sigma_{2,-} \sigma_{1,+})],\\
&=& [X_1 X_2,\nicefrac{1}{2}(\gamma(t) \lambda(t)^* \sigma_{2,-} \sigma_{1,+} - \gamma(t)^* \lambda(t) \sigma_{2,+} \sigma_{1,-})],\\
&=& \nicefrac{1}{2}\gamma(t) \lambda(t)^* [X_1 X_2,\sigma_{2,-}\sigma_{1,+}] -\nicefrac{1}{2} \gamma(t)^* \lambda(t) [X_1 X_2,\sigma_{2,+} \sigma_{1,-}].
\end{eqnarray*}
Combining yields,
\begin{eqnarray*}
\mathcal{L}_t(X_1X_2) &=& -\imath[X_1X_2,H(t)]+ \mathcal{L}_{L(t)}(X_1 X_2),\\
&=& \nicefrac{1}{2} \gamma(t)^* \lambda(t)[X_1,\sigma_{1,-}]\sigma_{2,+}X_2 + \nicefrac{1}{2} \gamma(t)^* \lambda(t) X_1[ \sigma_{2,+},X_2] \sigma_{1,-}\\
&&-\nicefrac{1}{2} \gamma(t)^* \lambda(t)(X_1 \sigma_{1,-}X_2 \sigma_{2,+}-\sigma_{1,-} X_1 \sigma_{2,+}X_2),\\
&&+\nicefrac{1}{2}|\lambda(t)|^2  (\sigma_{1,+} [X_1,\sigma_{1,-}] X_2 + [\sigma_{1,+},X_1] \sigma_{1,-} X_2]\\
&&+\nicefrac{1}{2} \lambda(t)^* \gamma(t) \sigma_{1,+} X_1 [X_2,\sigma_{2,-}] + \nicefrac{1}{2} \lambda(t)^* \gamma(t) [\sigma_{1,+},X_1] X_2 \sigma_{2,-}\\
&&\quad +\nicefrac{1}{2} \lambda(t)^* \gamma(t) (X_1 X_2 \sigma_{2,-}\sigma_{1,+}-\sigma_{1,+}\sigma_{2,-} X_1 X_2),\\
&&+\nicefrac{1}{2} |\gamma(t)|^2 (\sigma_{2,+} [X_2,\sigma_{2,-}]X_1 + X_1[\sigma_{2,+},X_2] \sigma_{2,-}),\\
&=& \gamma(t)^* \lambda(t) X_1 \sigma_{1,-} [\sigma_{2,+},X_2] +\nicefrac{1}{2} |\lambda(t)|^2(\sigma_{1,+}[X_1,\sigma_{1,-}]+[\sigma_{1,+},X_1]\sigma_{1,-})X_2\\
&& \quad + \lambda(t)^* \gamma(t) \sigma_{1,+} X_1[X_2,\sigma_{2,-}] +  \nicefrac{1}{2} |\gamma(t)|^2(\sigma_{2,+}[X_2,\sigma_{2,-}]+[\sigma_{2,+},X_2]\sigma_{2,-})X_1.
\end{eqnarray*}


\bibliographystyle{ieeetran}
\bibliography{rip,mjbib2004}

\begin{thebibliography}{10}
\providecommand{\url}[1]{#1}
\csname url@rmstyle\endcsname
\providecommand{\newblock}{\relax}
\providecommand{\bibinfo}[2]{#2}
\providecommand\BIBentrySTDinterwordspacing{\spaceskip=0pt\relax}
\providecommand\BIBentryALTinterwordstretchfactor{4}
\providecommand\BIBentryALTinterwordspacing{\spaceskip=\fontdimen2\font plus
\BIBentryALTinterwordstretchfactor\fontdimen3\font minus
  \fontdimen4\font\relax}
\providecommand\BIBforeignlanguage[2]{{%
\expandafter\ifx\csname l@#1\endcsname\relax
\typeout{** WARNING: IEEEtran.bst: No hyphenation pattern has been}%
\typeout{** loaded for the language `#1'. Using the pattern for}%
\typeout{** the default language instead.}%
\else
\language=\csname l@#1\endcsname
\fi
#2}}

\bibitem{DLCZ01}
L.-M. Duan, M.~D. Lukin, J.~I. Cirac, and P.~Zoller, ``Long-distance quantum
  communication with atomic ensembles and linear optics,'' \emph{Nature}, vol.
  414, pp. 413--418, 2001.

\bibitem{Kimb08}
H.~J. Kimble, ``The quantum internet,'' \emph{Nature}, vol. 453, pp.
  1023--1030, 2008.

\bibitem{CZKM97}
J.~I. Cirac, P.~Zoller, H.~J. Kimble, and H.~Mabuchi, ``Quantum state transfer
  and entanglement distribution among distant nodes in a quantum network,''
  \emph{Phys. Rev. Lett.}, vol.~78, no. 3221, 1997.

\bibitem{BvHJ07}
L.~Bouten, R.~{van Handel}, and M.~R. James, ``An introduction to quantum
  filtering,'' \emph{SIAM J. Control Optim.}, vol.~46, pp. 2199--2241, 2007.

\bibitem{HC93}
H.~Carmichael, \emph{An Open Systems Approach to Quantum Optics}.\hskip 1em
  plus 0.5em minus 0.4em\relax Berlin: Springer, 1993.

\bibitem{HRD09}
Q.~Y. He, M.~D. Reid, and P.~D. Drummond, ``Digital quantum memories with
  symmetric pulses,'' \emph{Opt. Express}, vol.~17, no.~12, pp. 9662--9668,
  2009.

\bibitem{DNSK12}
J.~Dilley, P.~{Nisbet-Jones}, B.~W. Shore, and A.~Kuhn, ``Single-photon
  absorption in coupled atom-cavity systems,'' \emph{Phys. Rev. A}, vol.~85,
  no. 023834, 2012.

\bibitem{ANK11}
A.~N. Korotkov, ``Flying microwave qubits with nearly perfect transfer
  efficiency,'' \emph{Phys. Rev. B}, vol.~84, no. 014510, 2011.

\bibitem{YCSEA13}
Y.~Yin, Y.~Chen, D.~Sank, P.~J.~J. O'Malley, T.~C. White, R.~Barends, J.~Kelly,
  E.~Lucero, M.~Mariantoni, A.~Megrant, C.~Neill, A.~Vainsencher, J.~Wenner,
  A.~N.~. Korotkov, A.~N. Cleland, and J.~M. Martinis, ``Catch and release of
  microwave photon states,'' \emph{Phys. Rev. Lett.}, vol. 110, no. 107001,
  2013.

\bibitem{YJ14}
N.~Yamamoto and M.~R. James, ``Zero-dynamics principle for perfect quantum
  memory in linear networks,'' \emph{New J. Phys.}, vol.~16, no. 073032, pp.
  1--30, 2014.

\bibitem{HP84}
R.~L. Hudson and K.~R. Parthasarathy, ``{Quantum Ito's formula and stochastic
  evolution},'' \emph{Commun. Math. Phys.}, vol.~93, pp. 301--323, 1984.

\bibitem{KRP92}
K.~Parthasarathy, \emph{An Introduction to Quantum Stochastic Calculus}.\hskip
  1em plus 0.5em minus 0.4em\relax Berlin: Birkhauser, 1992.

\bibitem{Mey95}
P.-A. Meyer, \emph{Quantum Probability for Probabilists}, 2nd~ed.\hskip 1em
  plus 0.5em minus 0.4em\relax Berlin-Heidelberg: Springer-Verlag, 1995.

\bibitem{BvH08}
L.~Bouten and R.~{van Handel}, ``On the separation principle of quantum
  control,'' in \emph{Quantum Stochastics and Information: Statistics,
  Filtering and Control (University of Nottingham, UK, 15 - 22 July 2006)},
  V.~P. Belavkin and M.~Guta, Eds.\hskip 1em plus 0.5em minus 0.4em\relax
  Singapore: World Scientific, 2008, pp. 206--238.

\bibitem{GJ09}
J.~Gough and M.~R. James, ``The series product and its application to quantum
  feedforward and feedback networks,'' \emph{IEEE Trans. Autom. Control},
  vol.~54, no.~11, pp. 2530--2544, 2009.

\bibitem{RL00}
R.~Loudon, \emph{The Quantum Theory of Light}, 3rd~ed.\hskip 1em plus 0.5em
  minus 0.4em\relax Oxford University Press, 2000.

\bibitem{GJM08}
G.~J. Milburn, ``Coherent control of single photon states,'' \emph{EPJ ST},
  vol. 159, no.~1, pp. 113--117, 2008.

\bibitem{GJN13}
J.~E. Gough, M.~R. James, and H.~I. Nurdin, ``Quantum filtering for systems
  driven by fields in single photon states and superposition of coherent states
  using non-markovian embeddings,'' \emph{Quantum Inf Process}, vol.~12, pp.
  1469--1499, 2013.

\bibitem{GJNC12}
J.~Gough, M.~R. James, H.~I. Nurdin, and J.~Combes, ``Quantum filtering for
  systems driven by fields in single-photon states or superposition of coherent
  states,'' \emph{Phys. Rev. A}, vol.~86, p. 043819, 2012.

\bibitem{GJN11}
J.~E. Gough, M.~R. James, and H.~I. Nurdin, ``Quantum master equation and
  filter for systems driven by fields in a single photon state,'' in
  \emph{Proceedings of the 50th IEEE Conference on Decision and Control (CDC)},
  2011, pp. 5570--5576.

\bibitem{PZJ16}
Y.~Pan, G.~Zhang, and M.~R. James, ``Analysis and control of quantum
  finite-level systems driven by single-photon input states,''
  \emph{Automatica}, vol.~69, pp. 18--23, 2016.

\bibitem{GJN10}
J.~E. Gough, M.~R. James, and H.~I. Nurdin, ``Squeezing components in linear
  quantum feedback networks,'' \emph{Phys. Rev. A}, vol.~81, pp. 023\,804--1--
  023\,804--15, 2010.

\bibitem{IYYYF11}
S.~Iida, M.~Yukawa, H.~Yonezawa, N.~Yamamoto, and A.~Furusawa, ``Experimental
  demonstration of coherent feedback control on optical field squeezing,''
  \emph{IEEE Trans. Autom. Control}, vol.~57, no.~8, pp. 2045--2050, 2012.

\end{thebibliography}

\end{document}